\documentclass{Interspeech}

\interspeechcameraready

\title{Crowdsourcing MUSHRA Tests in the Age of Generative Speech Technologies: A Comparative Analysis of Subjective and Objective Testing Methods}

\author[affiliation={1}]{Laura}{Lechler}
\author[affiliation={1}]{Chamran}{Moradi}
\author[affiliation={1}]{Ivana}{Balic}

\affiliation{}{Cisco Systems}{Europe}
\email{llechler@cisco.com, cmoradia@cisco.com, ibalic@cisco.com}

\usepackage{background}
\SetBgAngle{0}
\SetBgScale{1}
\SetBgColor{black}
\SetBgOpacity{1}
\SetBgPosition{0.66\paperwidth,-1.15\textheight}
\SetBgContents{
{\parbox{\paperwidth}{%
\footnotesize \textit{This is a preprint of a paper submitted to and accepted for INTERSPEECH 2025.}

\mbox{}\\
     \mbox{}\\}
     }%
}

\usepackage{cite}
\makeatletter
\def\@cite#1#2{\textnormal{[{#1\if@tempswa, #2\fi}]}} 
\makeatother

\begin{document}

\maketitle

\begin{abstract}
%
The MUSHRA framework is widely used for detecting subtle audio quality differences but traditionally relies on expert listeners in controlled environments, making it costly and impractical for model development. As a result, objective metrics are often used during development, with expert evaluations conducted later. While effective for traditional DSP codecs, these metrics often fail to reliably evaluate generative models. This paper proposes adaptations for conducting MUSHRA tests with non-expert, crowdsourced listeners, focusing on generative speech codecs. We validate our approach by comparing results from MTurk and Prolific crowdsourcing platforms with expert listener data, assessing test--retest reliability and alignment. Additionally, we evaluate six objective metrics, showing that traditional metrics undervalue generative models. Our findings reveal platform-specific biases and emphasize codec-aware metrics, offering guidance for scalable perceptual testing of speech codecs.

\end{abstract}

\section{Introduction}
Subjective perceptual studies are the gold standard for evaluating human perception of audio quality and related research tasks. Conducting these studies requires significant time and cost, often relying on specialized labs or resorting to small-sample internal listening tests with expert listeners. As a result, evaluation using objective metrics (e.g., PESQ \cite{rix_perceptual_2001}, POLQA \cite{beerends_perceptual_2013,beerends_perceptual_2013-1}, and NiSQA \cite{mittag_non-intrusive_2019,mittag_nisqa_2021}) is widely used. However, recent advancements in generative models have compromised their reliability, as many rely on ground-truth references. Generated outputs may meaningfully differ without perceptual degradation. Similarly, non-intrusive objective metrics often fail to correlate well with subjective ratings.

Many subjective test methodologies have successfully been adapted to a crowdsourcing environment, ensuring scalability and cost-effectiveness, compared to internal or professional laboratory tests. Examples are the application of MOS (mean opinion scores) for speech quality to the crowdsourcing environment \cite{ribeiro_crowdmos_2011}, leading to the implementation of ITU-T recommendation P.808 \cite{naderi_open_2020}, and the open-sourced adaptation of the Diagnostic Rhyme Test in multiple languages to assess speech intelligibility via crowdsourcing \cite{lechler_crowdsourced_2024}. The MUSHRA (multiple stimuli with hidden reference and anchor) framework of evaluating speech quality \cite{itu_method_2015} is a well-established test tool with high sensitivity to even small quality differences of high-quality speech codecs and other signal processing applications. 
 
 Studies investigating the effects of conducting MUSHRA tests with non-expert listeners demonstrate that relative rankings are largely consistent between the two participant groups \cite{schinkel-bielefeld_audio_2013,cartwright_fast_2016}, and some efforts have been made to facilitate online MUSHRA tests with the option to deploy these tests on crowdsourcing platforms \cite{cartwright_fast_2016,schoeffler_webmushra_2018,morrison2022reproducible}. 
 However, such studies note that absolute scores may differ between expert and non-expert listeners, and non-expert ratings appeared to be significantly above the expert ratings, indicating a ceiling effect for very high-quality conditions under test. 
 
 It is hence still a wide-spread assumption that the MUSHRA test framework is only suitable for an expert-listener audience (e.g., \cite{muller_speech_2024}). 
 This stance may be valid in certain application scenarios, such as evaluations for official accreditation and certification, where absolute scores are required that can be compared between test laboratories. However, for other use cases, such as continuous model evaluation throughout the development cycle and benchmarking against competitors, reliable relative rankings and scalability are priorities.

Due to the recent popularity of generative methods for speech coding, we revisit the question of whether crowdsourced MUSHRA tests can reliably evaluate such models. We propose adaptations to the MUSHRA protocol for use with non-expert listeners in a crowdsourced setting and release open-source tools for replication. Our approach enables faster and more scalable evaluations across diverse populations.

This work makes the following contributions: (i) it presents the first, to our knowledge, dedicated crowdsourced MUSHRA evaluation design, focusing on generative speech codecs, highlighting their unique perceptual challenges; (ii) it offers a direct comparison of two crowdsourcing platforms---Prolific and MTurk---revealing that while both replicate expert rankings with high test--retest reliability, only Prolific closely aligns with expert scores in absolute terms, whereas MTurk shows a ceiling effect; (iii) it provides a codec-specific analysis of subjective--objective metric correlation, showing that traditional metrics like PESQ and POLQA underestimate DNN-based codecs, while newer metrics such as SCOREQ show higher consistency across architectures. These insights inform evaluation design, highlight the value of subjective testing in generative codec research, and motivate further investigations into newer metrics.

\section{Crowdsourced MUSHRA testing}
\label{Crowdsourced-MUSHRA}

Traditional MUSHRA testing has always faced difficulties and high costs. Setting up controlled listening environments, recruiting trained participants, and managing equipment can be expensive and time-consuming. Moreover, ensuring consistent and reliable results can be challenging due to variations in listener focus and fatigue. By leveraging the power of the crowd, we can solve most of these difficulties. In addition, crowdsourced MUSHRA offers cost-efficiency, faster data collection, and greater reliability through a larger sample size. The larger and more diverse pool of participants accessible through crowdsourcing may be more representative of customer opinions and may avoid over-emphasizing the opinions of domain experts. Due to the low cost and fast collection pace, tests can be conducted more frequently.

Given the significant variability in response reliability among crowdworkers—stemming from factors like fatigue, impairments, carelessness, equipment differences, or malicious intent—rigorous participant screening is crucial. To address this, our tool enhances existing solutions (e.g., webMUSHRA \cite{schoeffler_webmushra_2018}) with two key features: real-time score screening to immediately discard unreliable responses based on deviation thresholds, and automatic test partitioning. This partitioning divides stimuli into smaller, configurable sub-tests for individual listeners, enabling evaluation of more conditions while mitigating fatigue and supporting diverse audio files.

Our procedure builds upon and adapts established guidelines from ITU-T P.808 \cite{itu_p808_2021}, which outlines best practices for crowdsourced MOS testing, and ITU-R BS.1534-3 \cite{itu_method_2015}, the standard for MUSHRA testing. We extend these recommendations to address practical challenges encountered when conducting MUSHRA tests with crowdsourced non-expert listeners.

\subsection{Pre-screening}
In line with subjective test recommendations for MOS tests in crowdsourced environments \cite{naderi_open_2020}, we applied quality filters on both MTurk and Prolific. Participants were required to have a success rate above 97\% and at least 100 completed tasks. 
Based on empirical observations, no further filters were used on MTurk, while on Prolific, we applied filters for first language ('English'), hearing difficulties ('no'), and cochlear implant ('no'). Depending on the platform, additional filters may be available; researchers should consider those relevant to their use case and report all filters used.

Whether native-speaker competence is required may depend on the research question. A study on cross-language evaluations of codec artefacts \cite{schinkel-bielefeld_is_2017} found no significant rating differences, though longer listening times were observed for foreign-language conditions in high-quality audio. However, native speakers are likely more able to judge intelligibility. 

\subsection{Qualification phase}
\label{sec:qualification}

Our survey flow included a questionnaire obtaining self-reported responses on several aspects relevant to participation eligibility, a digits-in-noise hearing test, and a training session similar to the actual rating task. The questionnaire is based on ITU recommendation P.800.1 \cite{itu_p800_1998} and assesses the listening device used, the participant’s tiredness level, the last participation in an audio listening test, and the self-reported level of hearing ability. Additionally, the participant’s gender, age, and level of English were recorded for statistical purposes. The hearing test consisted of six sets of three digits in noise from the P.808 toolkit \cite{naderi_open_2020}. Participants had to pass a threshold of correct digits. The training session consisted of one MUSHRA question with carefully selected conditions. Participants had three attempts to complete this correctly. If the validation criteria (scores cannot be equal to 0, reference must be ranked highest, the anchor must be ranked lowest) were not met, feedback was provided.

\subsection{Main rating task}
The main rating task contained up to three blocks of MUSHRA questions. Each block contained up to 26 test stimuli (including references and anchors). The number of MUSHRA questions per block therefore depended on the number of conditions directly compared, with fewer questions added with increasing numbers of conditions. 
If the validation criteria (cf. Subsection \ref{sec:qualification}) were also met in the main task, participants were allowed to complete up to three blocks. Allowing successful participants to rate several sections can reduce the error variance due to individual rating differences \cite{itu_p808_2021}. We recommend limiting the number of rating jobs to three at most to prevent fatigue. 

\subsection{Post-screening} 
The ITU Recommendation BS.1534-3 \cite{itu_method_2015}, tailored for expert-listener MUSHRA tests, describes procedures based on which certain listeners should be disqualified and outlier scores eliminated. 
We adapted these post-screening rules.
\paragraph*{Listener-level disqualification:} 
Listeners are disqualified entirely if they fail basic quality checks on more than 20\% of the MUSHRA questions in a test block. A failure occurs if either of the following conditions is met for a given question:
\begin{enumerate}
    \item The anchor is rated higher than the hidden reference, i.e. $r_{\text{anchor}} > r_{\text{ref}}$.
	\item All ratings for the non-anchor systems are identical, i.e. $\mathrm{Var}(r_1, r_2, \dots, r_K) = 0$, where $r_1, \dots, r_K$ are the scores for the $K$ conditions under test (excluding the anchor but including the hidden reference), and $\mathrm{Var}(\cdot)$ denotes variance.
\end{enumerate}
If 20\% of the questions in a test block is less than 1, we take $1$ as the threshold. 
\paragraph*{Score-level outlier removal:} After disqualifying certain listeners, we further proceed to removing outlying scores. A score is considered an outlier, and thus removed, if at least one of the above-mentioned criteria is met or it is identified as an outlier using the inter-quartile-range (IQR) rule explained in \cite{itu_method_2015}.

\subsection{Selecting test signals}
\label{signals}
We selected 40 clean speech wide-band test signals of high recording quality in English. 
A balanced sample of male, female, and children's speech was selected. Care was taken to choose full utterances of intelligible speech. As non-expert listeners were found to make less use of the looping functionality than expert listeners \cite{schinkel-bielefeld_audio_2013}, we suggest simplifying the task by not offering the looping functionality and reducing the audio duration. The listener can then focus on a manageable stretch of audio, while providing sufficient context. Although a duration of 10--12 seconds is recommended for expert listeners \cite{itu_method_2015}, we suggest using slightly shorter files for crowdsourcing. Our test signals had an average duration of 6.4 seconds (SD +/-- 1.3s). The vast participant pool of crowdsourcing platforms allows for testing more signals instead of longer signals.

\begin{figure}[!hbt]
  \centering
  \includegraphics[width=\linewidth]{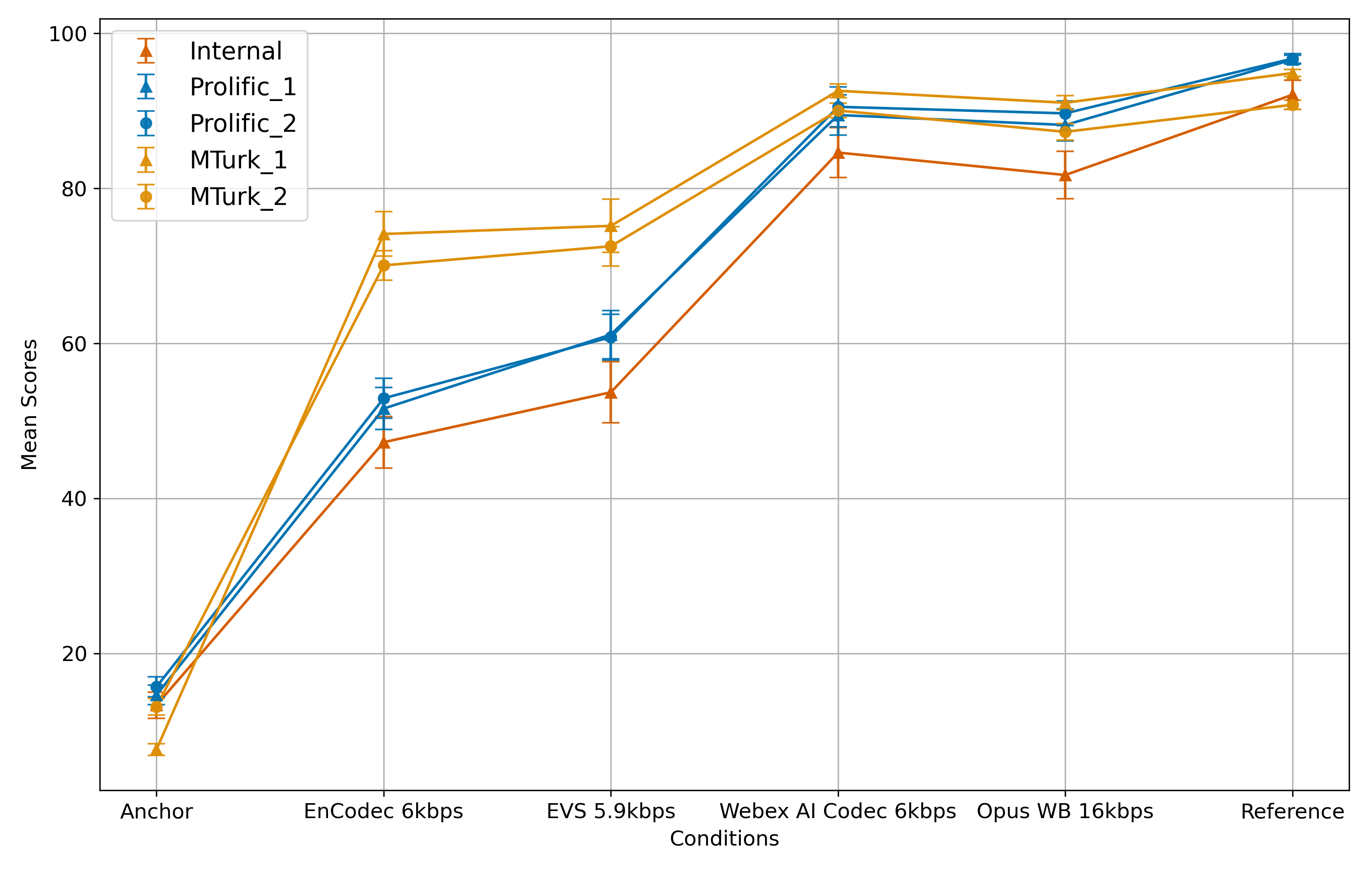}
  \caption{Repeatability and validity of crowdsourced test on two different crowdsourcing platforms.}
  \label{fig:repeat_1}
\end{figure}

\subsection{Choosing the anchor}

In MUSHRA tests, the anchor represents a clearly degraded version of the audio and defines the lowest point on the rating scale. It helps listeners calibrate their judgments by providing contrast to the high-quality reference. The type of degradation introduced in the anchor can strongly influence how participants use the scale and interpret differences between test conditions \cite{itu_method_2015}.

ITU-R BS.1534-3 recommends using an anchor that contains similar types of impairments as the systems being tested. This makes comparisons easier and reduces confusion, especially for non-expert listeners. Since we evaluated generative speech codecs, which typically introduce coding artifacts, we used Opus at 6 kbps as the anchor.

In a pilot experiment, we tried using a low-pass filtered version of the original signal—an anchor sometimes used in traditional tests—but participants found it harder to judge quality because the artifacts were of a different type (bandwidth limitation instead of coding distortion). This feedback confirmed that matching the anchor’s impairments to those of the test systems leads to more consistent and meaningful ratings.

\subsection{Aggregating results from different experiments}
For expert-listener tests, a maximum of 12 conditions is recommended to be evaluated in parallel, including the reference and all anchors \cite{itu_method_2015}. For crowdsourced tests with non-expert listeners, we assume this poses an unmanageable task. In our experience, a maximum of 6 conditions should not be exceeded in a crowdsourced test. However, further research into the attention span of crowdsourced non-expert listeners is required to establish the recommended maximum scientifically.

Due to this limitation, it is likely to split the conditions under test into several experiments. Care should be taken as to which systems are evaluated together, as context effects and a varying resolution of detail for very similar conditions are known challenges of this approach \cite{itu_method_2015}. Aggregating results from different experiments with the same anchor and reference requires renormalization, such that anchor and reference have the same values in all tests. We set the reference to 100 and the anchor to an average of the anchors from the three tests and re-normalized the other results accordingly. 


\section{Experiments}
\subsection{Experimental setup}
We set two main goals for evaluating our crowdsourced MUSHRA listening test. First, it should correlate well with an internal listening test conducted by expert listeners under controlled conditions (i.e., validity). Second, repeating the same test should yield the same results (i.e., repeatability). In this section, we examine both objectives through a study involving both internal and crowdsourced listening tests. We then compare the resulting subjective evaluations against several well-known objective metrics.

\begin{table}[th]
\footnotesize
  \caption{Pairwise correlations between MUSHRA tests}
  \centering
  \begin{tabular}{c c c c}
    \toprule
    Test 1 & Test 2 & Pearson & Spearman \\
    \midrule

Internal & Prolific 1 & 0.95 & 0.91 \\
Internal & Prolific 2 & 0.95 & 0.91 \\
Internal & MTurk 1 & 0.89 & 0.89 \\
Internal & MTurk 2 & 0.90 & 0.89 \\
\midrule
Prolific 1 & Prolific 2 & 0.98 & 0.95 \\
MTurk 1 & MTurk 2 & 0.99 & 0.94 \\

    \bottomrule
  \end{tabular}
\end{table}
\label{tab:subj_test_correlations}

We conducted listening tests using the 40 clean speech signals described in Section \ref{signals}. The test included four codecs: two AI-based codecs: Webex AI Codec (6 kbps) \cite{dhingra_webex_2024} and EnCodec (6 kbps) \cite{defossez2023high}, and two traditional DSP-based codecs: Opus (16 kbps) \cite{vos2013voice} and EVS (6 kbps) \cite{dietz_overview_2015}. Additionally, OPUS 6 kbps was used as an anchor. 
A MUSHRA test was run internally with in-house expert listeners in a quiet environment with wired headphones (4--6 votes per file). The same MUSHRA test was also run on two crowdsourcing platforms: MTurk and Prolific. The procedure for these tests was as described in Section \ref{Crowdsourced-MUSHRA}. We repeated the crowdsourced tests after several months to evaluate the repeatability of the results. 
For Prolific, we observed stable results with approximately 15 responses for each test item, while MTurk required around 25 responses per test item.

\subsection{Correlation testing}
In the following analyses, Pearson and Spearman correlations were calculated using per-condition mean scores (i.e., ratings for every item in each test condition). By examining the full, condition-level data, we capture how each platform’s ratings vary across all stimuli, providing a more fine-grained and robust assessment of their correlations.
%

%
\begin{figure*}[ht]
  \centering
  \includegraphics[width=0.92\textwidth]{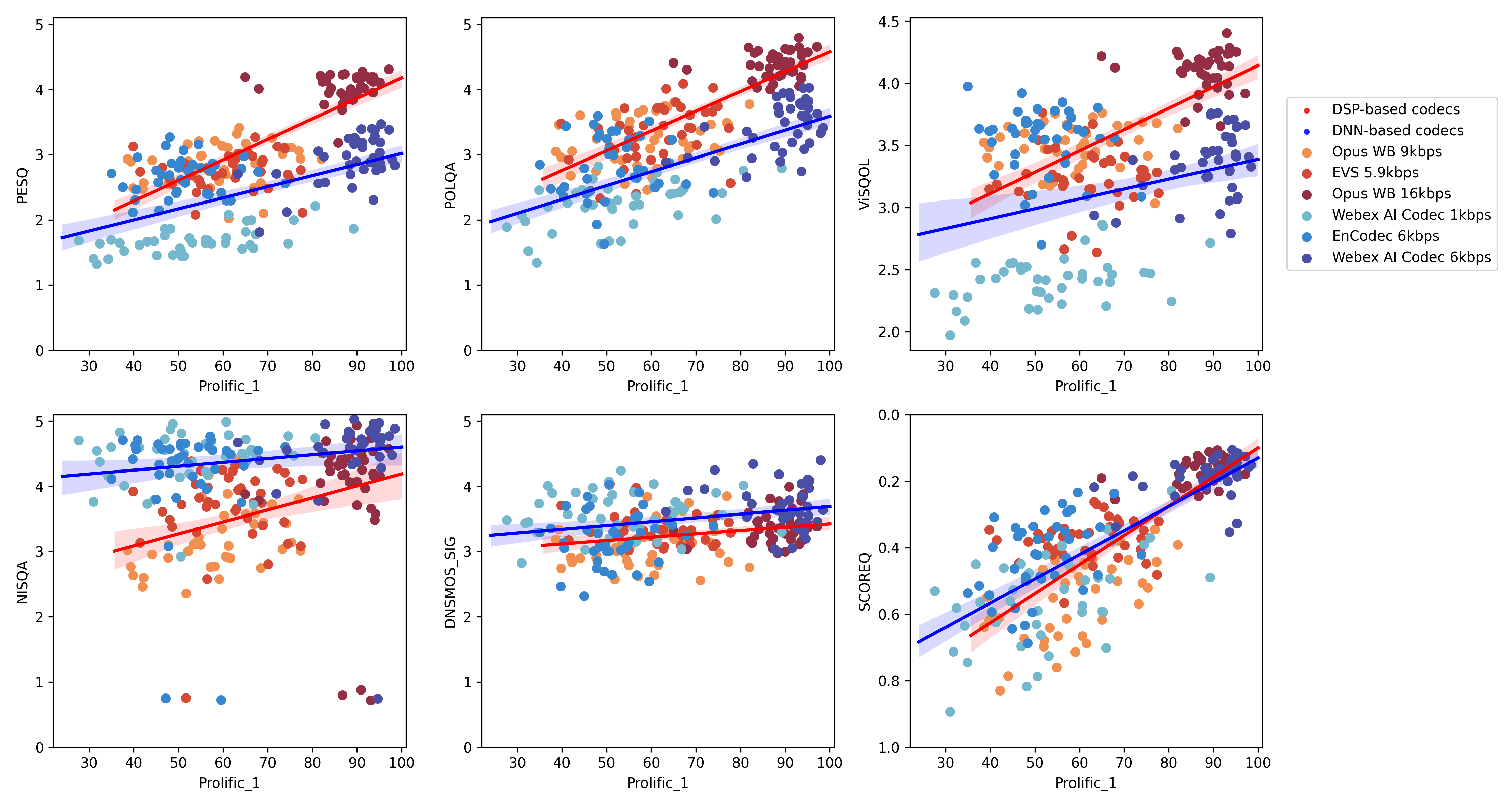}
  \caption{Linear regression plots for each objective metric against subjective scores (Prolific 1). Blue and red lines/dots correspond to 3 DNN- and 3 DSP-based codecs, respectively, across 40 test items.}
  \label{fig:MUSHRA_vs_objective}
\end{figure*}

\section{Results and discussion}
\subsection{Repeatability and Validity}
From Figure \ref{fig:repeat_1}, we see that both Prolific and MTurk listeners produce the same overall ranking of codecs as Internal listeners (i.e., “ground truth”). However, Prolific’s scores more closely trace the Internal curve, with closer absolute means and narrower confidence intervals, implying better alignment across all conditions. MTurk achieves the same ranking, but its mean ratings drift further from the Internal curve, particularly around mid-range codecs like EVS or Webex AI Codec.
Table 1 summarizes the Pearson and Spearman correlations between each crowdsourced platform’s scores and the Internal listening test. The high correlation values for both Prolific runs confirm that the per-file mean scores closely track the internal scores. By contrast, both MTurk tests maintain somewhat lower correlations, though still capturing the general ranking of conditions.



 We can also see in Figure 1 that both MTurk and Prolific exhibit strong internal consistency across repeated runs: Prolific 1 and Prolific 2 scores nearly coincide, while MTurk 1 and MTurk 2 show minor differences in the absolute overall scores. This reproducibility across independent trials underscores the overall reliability and robustness of these crowdsourced methods. The correlation test results reported in Table 1 confirm this test consistency for per-file mean scores.




\begin{table}[th]
\footnotesize
  \caption{Pearson correlations of Objective against Subjective Scores (Prolific 1) for all, DSP-, and DNN-based codecs.}
  
  \centering
  \begin{tabular}{c c c c }
    \toprule
    Obj. Metric & Overall & DSP Codecs & DNN Codecs \\
    \midrule

    PESQ &  0.69 &  0.79 &  0.58 \\
    POLQA &  0.72 &  0.78 &  0.69 \\
    ViSQOL &  0.55 &  0.69 &  0.30\\
    NISQA &  0.18 &  0.37 &  0.18\\
    DNSMOS-SIG &  0.21 &  0.28 &  0.29\\
    SCOREQ & --0.80 & --0.79 & --0.79\\

    \bottomrule
  \end{tabular}

\end{table}
\label{tab:obj_correlations}

\subsection{Comparison with Objective Metrics}
We compared our subjective test results with 6 objective metrics. Three of these metrics are well-known intrusive signal processing-based solutions: PESQ \cite{rix_perceptual_2001}, POLQA \cite{beerends_perceptual_2013,beerends_perceptual_2013-1}, and ViSQOL \cite{chinen_visqol_2020}. The remaining three metrics are DNN-based metrics: NiSQA \cite{mittag_non-intrusive_2019,mittag_nisqa_2021}, DNSMOS-SIG \cite{reddy2021dnsmos} (both non-intrusive) and SCOREQ \cite{ragano_scoreq_2024} (reference-based version).

Based on the strong correlation of Prolific tests with our internal results, the availability of scores for more conditions, and the higher participant numbers ensuring statistical robustness of file means, we decided to use the Prolific scores for the comparison of objective and subjective scores.
To better explore their performance on a variety of codecs, we added Opus (9 kbps) and Webex AI Codec (1 kbps) to the list of test conditions. 
By means of an informal listening test, we confirmed that the merged scores of a comparable test including these conditions were valid and well-correlated with internal assessments.

Figure \ref{fig:MUSHRA_vs_objective} presents six scatter plots, showing linear regression correlations between subjective scores and respective objective metrics for DSP-based (red) and DNN-based (blue) codecs. POLQA, PESQ, and ViSQOL correlated better with subjective scores for DSP-based codecs and systematically undervalued DNN-based codecs, evident in lower absolute scores and weaker correlations. NISQA and DNSMOS-SIG appear to underperform for both codec types, exhibiting poor alignment with the subjective results. By contrast, SCOREQ shows the strongest overall correlation with Prolific scores. The predictions remain consistently reliable across both DNN- and DSP-based codecs. We observe less scatter for higher-quality codecs, suggesting higher accuracy in predicting top-tier audio performance, regardless of codec architecture. Note that the y-axis for SCOREQ is reversed for clarity, since lower values indicate better quality for distance-based metrics.
Table 2 lists Pearson and Spearman correlations between objective and subjective scores for all (overall), only DSP-based, and only DNN-based codecs. These correlation results confirm the above findings. 

\section{Conclusion}

Our study demonstrates that crowdsourced MUSHRA tests—when designed with appropriate screening and procedural adjustments—can serve as reliable and repeatable alternatives to expert lab tests, even for generative speech codecs. We show that Prolific yields closer absolute alignment with expert ratings than MTurk, though both maintain reliable relative rankings. Among objective metrics, PESQ, POLQA, and ViSQOL correlated more strongly with subjective scores in DSP-based codecs but tended to underestimate DNN-based codecs. 
SCOREQ emerges as a promising tool for accurately tracking subjective speech quality across varying codec architectures. 
MUSHRA results, which may provide more stable results than MOS tests, serve as excellent reference data to further evaluate a metric's reliability and accuracy for a particular system under test.
In future work, we plan to leverage the proposed test design in an in-depth analysis of various metrics for neural audio codec evaluation.
Our test design and helper code to set up a MUSHRA test, including the qualification steps, can be found in our Github repository\footnote{https://github.com/cisco/multilingual-speech-testing}.



\bibliographystyle{IEEEtran}

\bibliography{references_final}
\end{document}